\documentclass[12pt]{iopart}

\begin{document}

\title[Spin-S Bose condensate in a magnetic field]{On microscopic theory of spin-$S$ Bose-Einstein condensate in a magnetic field}

\author{A.S. Peletminskii$^{1,2}$, S.V. Peletminskii$^{1}$, Yu.V. Slyusarenko$^{1}$}

\address{$^{1}$Akhiezer Institute for Theoretical Physics, National Science Center "Kharkov Institute of Physics
and Technology", 61108 Kharkov, Ukraine}
\address{$^{2}$The Abdus Salam International Centre for Theoretical
Physics, 34100 Trieste, Italy}
\ead{spelet@kipt.kharkov.ua}
\begin{abstract}

    The Bogoliubov model for weakly interacting Bose gas is extended to Bose-Einstein
    condensation (BEC) of spin-S atoms in a magnetic field. Equation for the
    vectorial order parameter valid at temperature $T\to 0$ is derived
    and its particular solution is found. This solution corresponds
    to the formation of BEC of atoms with a definite spin
    projection onto direction of a magnetic field. We study the thermodynamic stability
    of the found solution and obtain the expressions for low-lying collective modes.

\end{abstract}

\vspace{2pc} \noindent{\it
Keywords}: Bose-Einstein condensation, Spinor condensate, Weakly
interacting Bose gas, Quasiaverages, Thermodynamics, Excitation
spectra

\maketitle

\section{Introduction}
After the first remarkable experiments concerning the observation of
BEC in dilute gases of alkali atoms such as ${}^{87}$Rb
\cite{Anderson}, ${}^{23}$Na \cite{Davis}, and ${}^{7}$Li
\cite{Bradley} the interest to this phenomenon has revived
\cite{Pethick,Pitaevskii}. Later on, BEC has been also obtained in
other atomic species: atomic hydrogen \cite{Fried}, metastable
${}^{4}$He \cite{Pereira}, and ${}^{41}$K \cite{Modugno}. The
experimental realization of BEC has become possible due to the
progress of laser cooling and trapping techniques \cite{Chu}. The
carried out experiments have proved many predictions of the
microscopic theory for weakly interacting Bose gas, which originates
from the pioneering work of Bogoliubov \cite{Bogoliubov}.
Bogoliubov's theory has become almost the first theory in which it
was necessary to move essentially from the methods of standard
perturbative approach while describing the interaction effects.
However, this theory, in its original formulation, did not take into
account the internal degrees of freedom of atoms. The effect of spin
degrees of freedom for weakly interacting Bose gas (spinor BEC) has
been studied in \cite{Akhiezer}-\cite{Martikainen}.

The realization of optical trapping for atomic condensate
\cite{Stamper-Kurn} has stimulated theoretical interest to spinor
BEC. Bose condensation in a weakly interacting gas of bosonic atoms
has been studied theoretically by many authors both for spin-1
\cite{Ohmi}-\cite{Yip} and spin-2 \cite{Ueda1}, \cite{Martikainen}
bosons. These investigations are based on the effective interaction
Hamiltonians of two bosons, in which the interaction is
characterized by a definite number of interaction constants~--
$s$-wave scattering lengths. The number of scattering lengths is
determined by the total spin of two interacting bosons taking into
account the symmetry properties of their wave function. For example,
in case of spin-1 atoms the interaction Hamiltonian contains two
interaction constants \cite{Ohmi}-\cite{Yip}, in case of spin-2
atoms there are three interaction constants \cite{Ueda1},
\cite{Martikainen}. Thus, as the spin value of atoms grows, the
number of constants, which characterize the interaction of two
bosons, is increased. Note that in the mentioned effective
Hamiltonians it is difficult to interpret the physical nature of the
separated term of non-relativistic interaction not associated with
neither potential nor spin-exchange interactions (see e.g.
\cite{Ueda1}).

In this paper we study a weakly interacting Bose gas of particles
with arbitrary integer spin $S$ in a magnetic field (see also
\cite{Akhiezer}). We start from the microscopic interaction
Hamiltonian for two spin-$S$ bosons. This Hamiltonian is specified
by two functions, which describe potential and spin-exchange
interactions of spin-$S$ atoms. According to general rules of
quantum mechanics, we pass from the pairwise interaction of two
bosons to the standard expression for binary interaction of
arbitrary number of bosons in the second quantization
representation. By solving the multichannel scattering problem for
the considered Hamiltonian we could find, in principle, all
scattering lengths in terms of the functions characterizing the
potential and spin-exchange interactions. Thereby, it would be
possible to obtain the Hamiltonians analogous to the above mentioned
effective interaction Hamiltonians (see e.g. \cite{Ueda1}). However,
the use of the microscopic Hamiltonian enables to restrict ourself
by two interaction constants even in the case of arbitrary spin when
studying the ground state, stability, and excitations in a weakly
interacting gas in the presence of BEC.

\section{Method of quasiaverages and the model with a separated condensate}

To describe the system with a spontaneously broken symmetry we
address to the method of quasiaverages
\cite{Bogoliubov0,Bogoliubov1}. According to this method the Gibbs
statistical operator is modified so that it possesses the symmetry
of degenerate state. This modification is usually done by
introducing the infinitesimal "source" $\nu\hat{F}$ ($\nu\to 0$)
into the Gibbs exponent, which has the symmetry of phase under
consideration. Then, the average value of any physical quantity
$\hat{A}$ is defined as
\begin{equation}
\prec\hat{A}\succ=\lim_{\nu\rightarrow 0}\lim_{V\rightarrow\infty}
{\rm Tr}\,\hat{w}_{\nu}\hat{A}, \label{2.2}
\end{equation}
where the Gibbs statistical operator $\hat{w}_{\nu}$ has the form
\begin{equation}
\hat{w}_{\nu}=\exp{\left(\Omega_{\nu}-\beta(\hat{H}-\mu\hat{N}+
\nu\hat{F})\right)}. \label{2.3}
\end{equation}
Here $\beta=1/T$, $\mu$ are the reciprocal temperature and
chemical potential respectively and $\hat{H}$, $\hat{N}$ are the
system Hamiltonian and the particle number operator. The
thermodynamic potential $\Omega_{\nu}$ being a function of
thermodynamic parameters $\beta$, $\mu$ is found from the normalization condition
${\rm Tr} \,\hat{w}_{\nu}=1$. Notice that the limits in
(\ref{2.2}) are not permutable.

Consider a gas of condensed bosonic atoms with spin $S$. The
formation of a condensate is accompanied by the gauge symmetry
breaking and, therefore, in order to remove this kind of degeneracy
we should choose the "source" $\nu\hat{F}$ in (\ref{2.3}) such that
$[\hat{w}_{\nu},\hat{N}]\neq 0 $ ($\hat{N}$ is the generator of
phase transformation),
\begin{equation}
\nu\hat{F}=\nu_{\alpha}\int d^{3}x(\hat{\psi}^{\dag}_{\alpha}({\bf
x})+\hat{\psi}_{\alpha}({\bf x})), \label{2.4}
\end{equation}
where $\hat{\psi}^{\dag}_{\alpha}({\bf x})$,
$\hat{\psi}_{\alpha}({\bf x})$ are the creation and annihilation
operators with index $\alpha$ taking $2S+1$ values (the summation
over repeated indices is assumed). Then, according to (\ref{2.2}),
(\ref{2.3}), $\prec\hat{\psi}_{\alpha}({\bf
x})\succ=V^{-1/2}\prec\hat{a}_{0\alpha}\succ\sim 1$ that corresponds
to the formation of atomic condensate with momenta ${\bf p}=0$. The
order parameter $\Psi_{\alpha}=V^{-1/2}\prec\hat{a}_{0\alpha}\succ$
is called the condensate wave function.

The method of quasiaverages and the spatial correlation decay
principle enable to justify the replacement of creation and
annihilation operators of atoms with momentum ${\bf p}=0$ by
$c$--numbers,
$\hat{a}_{0\alpha},\,\hat{a}_{0\alpha}^{\dag}\rightarrow
\sqrt{V}\Psi_{\alpha},\,\sqrt{V}\Psi^{*}_{\alpha}$
\cite{Bogoliubov0}-\cite{Peletminskii} (the condensate separation
procedure).

The basic statement of the method of quasiaverages applied
to the description of BEC consists in the following
\cite{Bogoliubov0}-\cite{Peletminskii}: the Gibbs statistical
operator is replaced by
\begin{equation}
\hat{w}(\Psi)=\exp{(\Omega(\Psi)-\beta(\hat{H}(\Psi)-
\mu\hat{N}(\Psi)))}, \label{2.6}
\end{equation}
where $\Psi=\{\Psi_{\alpha}, \Psi_{\alpha}^{*}\}$ is found from the
following equation:
\begin{equation}
{\partial\Omega(\Psi)\over\partial\Psi}=0. \label{2.7}
\end{equation}

\section{The ground state of spin-$S$ condensate in a
magnetic field}

In this section we study one of the possible ground states of
spin-$S$ BEC in a magnetic field. In doing so, we start from the
Hamiltonian $\hat{\cal H}=\hat{H}-\mu\hat{N}$, which determines the
Gibbs statistical operator (\ref{2.6}) and has the following form:
\begin{equation}
\hat{\cal H}=\hat{\cal H}_{0}+\hat{\cal H}_{\rm p}+\hat{\cal
H}_{\rm e}, \label{3.1}
\end{equation}
where
\begin{eqnarray}
\hat{\cal H}_{0}=\sum_{\bf p}\hat{a}^{\dag}_{{\bf
p}\alpha}\left[\left(\varepsilon_{p}-\mu\right)\delta_{\alpha\beta}-{\bf
h}{\bf S}_{\alpha\beta}\right]\hat{a}_{{\bf p}\beta}, \quad
\varepsilon_{p}={p^{2}\over
2M}\label{3.2} \\
\hat{\cal H}_{\rm p}={1\over 2V}\sum_{{\bf p}_{1}\ldots{\bf
p}_{4}}U({\bf p}_{13})\delta_{{\bf p}_{1}+{\bf p}_{2},{\bf
p}_{3}+{\bf p}_{4}}\hat{a}^{\dag}_{{\bf
p}_{1}\alpha}\hat{a}^{\dag}_{{\bf p}_{2}\beta}\hat{a}_{{\bf
p}_{3}\alpha}\hat{a}_{{\bf p}_{4}\beta}, \label{3.3}\\
\hat{\cal H}_{\rm e}={1\over 2V}\sum_{{\bf p}_{1}\ldots{\bf
p}_{4}}J({\bf p}_{13})\delta_{{\bf p}_{1}+{\bf p}_{2},{\bf
p}_{3}+{\bf p}_{4}}\hat{a}^{\dag}_{{\bf
p}_{1}\alpha}\hat{a}^{\dag}_{{\bf p}_{2}\beta}{\bf
S}_{\alpha\gamma}{\bf S}_{\beta\rho}\hat{a}_{{\bf
p}_{3}\gamma}\hat{a}_{{\bf p}_{4}\rho}. \label{3.4}
\end{eqnarray}
Here ${\bf S}_{\alpha\beta}$ are the spin matrices, $U({\bf
p}_{13})$, $J({\bf p}_{13})$ (${\bf p}_{13}={\bf p}_{1}-{\bf
p}_{3}$) are the Fourier transforms of the amplitudes of potential
and spin-exchange interactions respectively, and ${\bf h}={g{\bf
H}/S}$ ($g$ is the Bohr magneton, and ${\bf H}$ is an external
magnetic field). For our next calculations it is convenient to
introduce the so-called ladder operators
$\hat{S}_{\pm}=\hat{S}_{x}\pm i\hat{S}_{y}$. Then, their nonzero
matrix elements in the representation, where $\hat{S}_{z}$ is a
diagonal matrix,
$\langle\alpha|\hat{S}_{z}|\alpha^{\prime}\rangle=\alpha\delta_{\alpha\alpha^{\prime}}$,
have the form
\begin{eqnarray}
\langle\alpha+1|\hat{S}_{+}|\alpha\rangle=\sqrt{S(S+1)-\alpha(\alpha+1)}, \label{3.5} \\
\langle\alpha-1|\hat{S}_{-}|\alpha\rangle=\sqrt{S(S+1)-\alpha(\alpha-1)}.
\nonumber
\end{eqnarray}

Now we separate the ${\bf p}=0$ components $\hat{a}_{0\alpha}$ in
the Hamiltonian (the replacement of $\hat{a}_{0\alpha}$ by
$c$~--numbers, $\hat{a}_{0\alpha}\to \sqrt{V}\Psi_{\alpha}$) and
keep the terms only up to second order in $\hat{a}_{{\bf
p}\alpha}$. We omit the higher order terms, since they should be
taken into account only when examining the interaction between
quasiparticles, which we will introduce in the next section. As a
result the Hamiltonian takes the form $\hat{\cal H}\approx{\cal
H}^{(0)}+\hat{\cal H}^{(2)}$. The explicit expression for
${\cal H}^{(0)}$, which contains only $c$~--numbers
$\Psi_{\alpha}$ reads
\begin{equation}
{{\cal H}^{(0)}\over V}={U(0)\over2}(\Psi^{*}\Psi)^{2}+{J(0)\over
2}(\Psi^{*}\hat{\bf S}\Psi)^{2}-{\bf h}\Psi^{*}\hat{\bf
S}\Psi-\mu\Psi^{*}\Psi, \label{3.6}
\end{equation}
where
\begin{equation} \label{3.6'}
\Psi^{*}\Psi=\Psi_{\alpha}^{*}\Psi_{\alpha}, \quad
\Psi^{*}\hat{\bf S}\Psi=\Psi_{\alpha}^{*}{\bf
S}_{\alpha\beta}\Psi_{\beta}
\end{equation}
The explicit form for $\hat{\cal H}^{(2)}$ will be written in the
next section.

Next, making use the normalization condition ${\rm Tr}\,\hat{w}=1$,
we find immediately the thermodynamic potential density
$\omega=\Omega T/V$ in the leading approximation (neglect of
quasiparticles; $T\to 0$) of the model for weakly interacting Bose
gas,
\begin{equation}
\omega={U(0)\over 2}(\Psi^{*}\Psi)^{2}+{J(0)\over
2}(\Psi^{*}\hat{\bf S}\Psi)^{2}-{\bf h}\Psi^{*}\hat{\bf
S}\Psi-\mu\Psi^{*}\Psi. \label{3.7}
\end{equation}
Therefore, Eq. (\ref{2.7}) for $\Psi_{\alpha}$ takes the form
\[
\mu\Psi_{\alpha}-U(0)(\Psi^{*}\Psi)\Psi_{\alpha}-J(0)(\Psi^{*}\hat{\bf
S}\Psi){\bf S}_{\alpha\beta}\Psi_{\beta}+{\bf h}{\bf
S}_{\alpha\beta}\Psi_{\beta}=0.
\]
If to introduce the normalized spin functions $\zeta_{\alpha}$,
$\Psi_{\alpha}=\sqrt{n}\zeta_{\alpha}$, where
$n=\Psi_{\alpha}\Psi^{*}_{\alpha}$ is the condensate density and
$\zeta_{\alpha}\zeta_{\alpha}^{*}=1$, then the latter equation is
written as
\begin{equation}
\mu\zeta_{\alpha}-nU(0)\zeta_{\alpha}-nJ(0)(\zeta^{*}\hat{\bf
S}\zeta){\bf S}_{\alpha\beta}\zeta_{\beta}+{\bf h}{\bf
S}_{\alpha\beta}\zeta_{\beta}=0 \label{3.8}.
\end{equation}
Assuming the vector ${\bf h}$ directed along $z$-axis (${\bf
h}=(0,0,h)$), its solution $\zeta_{\alpha}^{(m)}$ being an
eigenfunction of $\hat{S}_{z}$,
$(\hat{S}_{z})_{\alpha\beta}\zeta_{\beta}^{(m)}=m\zeta_{\alpha}^{(m)}$,
has the form
\begin{equation}
\zeta_{\alpha}^{(m)}=\delta_{\alpha m} \label{3.9}.
\end{equation}
Next, taking into account that $\hat{S}_{z}$ is a
diagonal matrix, whereas $\hat{S}_{\pm}$ have no
diagonal matrix elements in the considered representation of spin
matrices, one finds from Eq. (\ref{3.8})
\begin{equation}
n={{\mu+mh}\over{U(0)+m^{2}J(0)}} \label{3.10}.
\end{equation}
The obtained formulae (\ref{3.9}), (\ref{3.10}) result in the
following expression for the thermodynamic potential density:
\begin{equation}
\omega=-{1\over 2}{{(\mu+mh)^{2}}\over{U(0)+m^{2}J(0)}}
\label{3.11}.
\end{equation}

We are now in a position to study the stability of possible ground
states (\ref{3.9}). In the considered approximation, the
thermodynamic potential of the normal state is zero (the order
parameter $\Psi_{\alpha}$ vanishes). Therefore, for the stability of
the ground state under consideration, the density of thermodynamic
potential must be negative, $\omega<0$ and, consequently, according
to (\ref{3.11}), we can write the necessary condition of
thermodynamic stability,
\begin{equation}
U(0)+m^{2}J(0)>0 \label{3.12}.
\end{equation}
Let us find now such spin projections $m$, which correspond to the
minimum of potential (\ref{3.11}). For simplicity, we study the case
of $h=0$ (or sufficiently weak $h$). Then,
\begin{equation*}
\omega=-{\mu^{2}\over 2}{1\over{U(0)+m^{2}J(0)}}<0. \label{3.13}
\end{equation*}
As it can be easily seen that in contrast to usual Bogoliubov's
theory, in which $U(0)>0$ (the necessary condition of stability),
the negative values of $U(0)$ are also permissible. Therefore, we
have the following three situations:

1) $U(0)>0$, $J(0)>0$. In this case the requirement (\ref{3.12}) is
automatically satisfied. The density of thermodynamic potential
(\ref{3.11}) has a minimum at $m=0$ in which
$\omega=-\mu^{2}/2U(0)$. We call this case as antiferromagnetic
ordering.

2) $U(0)>0$, $J(0)<0$ but such that the requirement (\ref{3.12})
should be satisfied. The minimum of $\omega$ is reached for $m_{\rm
min}=\pm(m_{\rm c}-1)$, where
\begin{equation}
m_{\rm c}=\left[\left(-{U(0)\over J(0)}\right)\right]^{1/2}, \quad
m_{\rm c}\leq S+1 \label{3.14}
\end{equation}
(the square brackets denote an integer part). This case
corresponds to ferromagnetic ordering.

3) $U(0)<0$, $J(0)>0$ but again, such that $U(0)+m^{2}J(0)>0$. Here
the minimum of $\omega$ is given by the spin projections $m_{\rm
min}=\pm(m_{\rm c}+1)$, where $m_{\rm c}$ is also defined by
(\ref{3.14}) but with $U(0)<0$, $J(0)>0$. This case also corresponds
to ferromagnetic ordering.

\section{Low-lying collective modes}

In this section we obtain the excitation spectra of spin-$S$ BEC by
employing the well-known diagonalization procedure (Bogoliubov's
$u-v$ transformations \cite{Bogoliubov}) for the Hamiltonian
quadratic in creation and annihilation operators. Note that the
excitation spectra can also be found as a result of the
linearization of the Gross-Pitaevskii equation \cite{Gross,Pit} for
the condensate wave function (see e.g. \cite{Ohmi}).

The part of the spin-exchange interaction Hamiltonian (\ref{3.4}),
which is quadratic in $\hat{a}_{{\bf p}\alpha}$, (${\bf p}\neq 0$),
has the form
\begin{equation}
\fl \hat{\cal H}_{\rm e}^{(2)}=J(0)\Psi^{*}\hat{\bf
S}\Psi\sum_{\bf p}\hat{a}^{\dagger}_{{\bf p}}\hat{\bf
S}\hat{a}_{{\bf p}} +{1\over 2}\sum_{{\bf p}}J({\bf
p})\left[(\hat{a}^{\dagger}_{{\bf p}}\hat{\bf
S}\Psi)(\hat{a}^{\dagger}_{-{\bf p}}\hat{\bf
S}\Psi)+(\hat{a}^{\dagger}_{{\bf p}}\hat{\bf
S}\Psi)(\Psi^{*}\hat{\bf S}\hat{a}_{{\bf p}})+{\rm h.c.}\right],
\label{4.1}
\end{equation}
where we have used the notations (\ref{3.6'}). Taking into account that
$\hat{S}_{x}={1\over 2}(\hat{S}_{+}+\hat{S}_{-})$,
$\hat{S}_{y}=-{i\over 2}(\hat{S}_{+}-\hat{S}_{-})$ and bearing in mind
(\ref{3.5}) for non-zero matrix
elements of $\hat{S}_{\pm}$ as well as the explicit form of
the condensate wave function $\Psi_{\alpha}^{(m)}=\sqrt{n}\delta_{\alpha m}$,
one gets
\begin{eqnarray*}
\fl (\Psi^{*}\hat{\bf S}\Psi)(\hat{a}_{\bf p}^{\dagger}\hat{\bf
S}\hat{a}_{\bf p})=nm\sum_{\alpha}\alpha\hat{a}^{\dagger}_{{\bf
p}\alpha}\hat{a}_{{\bf p}\alpha}, \\
\fl (\hat{a}^{\dagger}_{{\bf p}}\hat{\bf
S}\Psi)(\hat{a}^{\dagger}_{-{\bf p}}\hat{\bf
S}\Psi)=nm^{2}\hat{a}^{\dagger}_{{\bf p}m}\hat{a}^{\dagger}_{-{\bf
p}m}+{n\over 2}S_{m}S_{-m}\left(\hat{a}^{\dagger}_{{\bf
p}m-1}\hat{a}^{\dagger}_{-{\bf p}m+1}+\hat{a}^{\dagger}_{{\bf
p}m+1}\hat{a}^{\dagger}_{-{\bf p}m-1}\right),\\
\fl (\hat{a}^{\dagger}_{{\bf p}}\hat{\bf S}\Psi)(\Psi^{*}\hat{\bf
S}\hat{a}_{\bf p})=nm^{2}\hat{a}^{\dagger}_{{\bf p}m}\hat{a}_{{\bf
p}m}+{n\over 2}\left(S^{2}_{-m}\hat{a}^{\dagger}_{{\bf
p}m-1}\hat{a}_{{\bf p}m-1}+S^{2}_{m}\hat{a}^{\dagger}_{{\bf
p}m+1}\hat{a}_{{\bf p}m+1}\right).
\end{eqnarray*}
where the following notation has been introduced:
\begin{equation*}
S_{m}=\sqrt{S(S+1)-m(m+1)}. \label{4.2}
\end{equation*}
Hence, $\hat{\cal H}_{\rm e}^{(2)}$ takes the form
\begin{eqnarray*}
\fl \hat{\cal H}_{\rm e}^{(2)}=J(0)nm\sum_{\bf
p}\left[(m-1)\hat{a}^{\dagger}_{{\bf p}m-1}\hat{a}_{{\bf
p}m-1}+m\hat{a}^{\dagger}_{{\bf p}m}\hat{a}_{{\bf
p}m}+(m+1)\hat{a}^{\dagger}_{{\bf p}m+1}\hat{a}_{{\bf
p}m+1}\right] \\
+J(0)nm\sum_{{\bf p},\alpha}\alpha\hat{a}^{\dagger}_{{\bf
p}\alpha}\hat{a}_{{\bf p}\alpha}+{n\over 2}\sum_{\bf p}J({\bf
p})m^{2}\left[\hat{a}^{\dagger}_{{\bf p}m}\hat{a}^{\dagger}_{-{\bf
p}m}+2\hat{a}^{\dagger}_{{\bf p}m}\hat{a}_{{\bf
p}m}+\hat{a}_{-{\bf p}m}\hat{a}_{{\bf p}m}\right] \\
+{n\over 2}\sum_{{\bf p}}J({\bf
p})\left[S_{m}S_{-m}(\hat{a}^{\dagger}_{{\bf
p}m-1}\hat{a}^{\dagger}_{-{\bf p}m+1}+\hat{a}_{-{\bf
p}m+1}\hat{a}_{{\bf p}m-1})\right] \\
+{n\over 2}\sum_{{\bf p}}J({\bf p})
\left[S^{2}_{-m}\hat{a}^{\dagger}_{{\bf p}m-1}\hat{a}_{{\bf
p}m-1}+S^{2}_{m}\hat{a}^{\dagger}_{{\bf p}m+1}\hat{a}_{{\bf
p}m+1}\right], \quad \alpha\neq m-1, \,m,\,m+1.
\end{eqnarray*}
In this formula, the summation index $\alpha$ in the second term
takes all values of spin projections except $m-1$, $m$, and $m+1$
(these three projections we have separated off and written them as
the first term in $\hat{\cal H}_{\rm e}^{(2)}$). In a similar
manner, keeping the terms only of second order in $\hat{a}_{{\bf
p}\alpha}$, one finds according to (\ref{3.2}), (\ref{3.3})
\begin{eqnarray}
\fl \hat{\cal H}_{0}^{(2)}=\sum_{\bf
p}\left[\left(\varepsilon_{p}-\mu\right)(\hat{a}^{\dagger}_{{\bf
p}m-1}\hat{a}_{{\bf p}m-1}+\hat{a}^{\dagger}_{{\bf
p}m}\hat{a}_{{\bf p}m}+\hat{a}^{\dagger}_{{\bf p}m+1}\hat{a}_{{\bf
p}m+1})\right] \nonumber \\
-h\sum_{{\bf p}}\left[(m-1)\hat{a}^{\dagger}_{{\bf
p}m-1}\hat{a}_{{\bf p}m-1}+m\hat{a}^{\dagger}_{{\bf
p}m}\hat{a}_{{\bf p}m}+(m+1)\hat{a}^{\dagger}_{{\bf
p}m+1}\hat{a}_{{\bf p}m+1}\right] \nonumber \\ +\sum_{{\bf
p},\alpha}\left(\varepsilon_{p}-\mu-\alpha
h\right)\hat{a}^{\dagger}_{{\bf p}\alpha}\hat{a}_{{\bf p}\alpha} ,
\quad \alpha\neq m-1,\,m,\,m+1, \label{4.3}
\end{eqnarray}
and
\begin{eqnarray}
\fl \hat{\cal H}^{(2)}_{\rm p}=U(0)n\sum_{\bf
p}\left[\hat{a}^{\dagger}_{{\bf p}m-1}\hat{a}_{{\bf
p}m-1}+\hat{a}^{\dagger}_{{\bf p}m}\hat{a}_{{\bf
p}m}+\hat{a}^{\dagger}_{{\bf p}m+1}\hat{a}_{{\bf
p}m+1}\right]+U(0)n\sum_{{\bf p},\alpha}\hat{a}^{\dagger}_{{\bf
p}\alpha}\hat{a}_{{\bf
p}\alpha} \label{4.4} \\
+{n\over 2}\sum_{\bf p}U({\bf p})\left[\hat{a}^{\dagger}_{{\bf
p}m}\hat{a}^{\dagger}_{-{\bf p}m}+2\hat{a}^{\dagger}_{{\bf
p}m}\hat{a}_{{\bf p}m}+\hat{a}_{-{\bf p}m}\hat{a}_{{\bf
p}m}\right], \quad \alpha\neq m-1,\,m,\,m+1. \nonumber
\end{eqnarray}
When obtaining (\ref{4.3}) we have employed the fact that ${\bf h}$
is directed along $z$-axis, ${\bf h}=(0,0,h)$. Next, using Eq.
(\ref{3.10}) to eliminate the chemical potential $\mu$ in
(\ref{4.3}), we recast the total Hamiltonian $\hat{\cal
H}^{(2)}=\hat{\cal H}^{(2)}_{0}+\hat{\cal H}^{(2)}_{\rm p}+\hat{\cal
H}^{(2)}_{\rm e}$ that is quadratic in creation and annihilation
operators in the following form:
\begin{equation}
\hat{\cal H}^{(2)}=\hat{\cal H}_{\alpha}^{(2)}+\hat{\cal
H}^{(2)}(m)+\hat{\cal H}^{2}(m-1,m+1), \label{4.5}
\end{equation}
where
\begin{equation}
\fl \hat{\cal H}_{\alpha}^{(2)}=\sum_{{\bf
p},\alpha}\left[\varepsilon_{p}-J(0)nm(m-\alpha)+h(m-\alpha)\right]
\hat{a}^{\dagger}_{{\bf p}\alpha}\hat{a}_{{\bf p}\alpha}, \quad
\alpha\neq m-1,\,m,\,m+1, \label{4.6}
\end{equation}
\begin{equation}
\fl \hat{\cal H}^{(2)}(m)=\sum_{\bf
p}\left[\varepsilon_{p}+g_{m}({\bf
p})\right]\hat{a}^{\dagger}_{{\bf p}m}\hat{a}_{{\bf p}m}+{1\over
2}\sum_{\bf p}g_{m}({\bf p})\left[\hat{a}^{\dagger}_{{\bf
p}m}\hat{a}^{\dagger}_{-{\bf p}m}+\hat{a}_{{\bf p}m}\hat{a}_{-{\bf
p}m}\right], \label{4.7}
\end{equation}
\begin{eqnarray}
\fl \hat{\cal H}^{(2)}(m-1,m+1)=\sum_{\bf
p}\left[\varepsilon_{p}-h+\beta_{m}(\bf
p)\right]\hat{a}^{\dagger}_{{\bf p}m+1}\hat{a}_{{\bf p}m+1} \nonumber \\
+\sum_{\bf p}\left[\varepsilon_{p}+h+\beta_{-m}(\bf
p)\right]\hat{a}^{\dagger}_{{\bf p}m-1}\hat{a}_{{\bf
p}m-1} \nonumber \\
+\sum_{\bf p}\alpha_{m}({\bf p})\left[\hat{a}^{\dagger}_{{\bf
p}m-1}\hat{a}^{\dagger}_{-{\bf p}m+1}+\hat{a}_{{\bf
p}m-1}\hat{a}_{-{\bf p}m+1}\right] \label{4.8}.
\end{eqnarray}
The introduced quantities $\alpha_{m}({\bf p})$, $\beta_{m}({\bf
p})$, and $g_{m}({\bf p})$ are given by
\begin{eqnarray}
\alpha_{m}({\bf p})={n\over 2}J({\bf p})S_{m}S_{-m}, \label{4.9} \\
\beta_{m}({\bf p})={n\over 2}J({\bf p})S^{2}_{m}+nJ(0)m, \label{4.10} \\
g_{m}({\bf p})=n(U({\bf p})+m^{2}J({\bf p})).  \label{4.11}
\end{eqnarray}

Now we are in a position to carry out the diagonalization procedure
of the total Hamiltonian (\ref{4.5}) quadratic in creation and
annihilation operators. In this connection we note that the
"Hamiltonians" (\ref{4.6})-(\ref{4.8}) contain the creation and
annihilation operators with not overlapping sets of indices
$\alpha$, $m$, $m-1$, $m+1$ ($\alpha\neq m-1,\,m,\,m+1$). Therefore,
we can perform their diagonalization independently. The evidence of
this statement also follows from the fact that (\ref{4.5}) can be
considered as the Hamiltonian of the system consisting of four kinds
($m,\,m\pm 1,\,\alpha$) of noninteracting particles.

The "Hamiltonian" $\hat{\cal H}_{\alpha}^{(2)}$ has already a diagonal form
with the following spectrum:
\begin{equation}
\omega_{m,\alpha}({\bf p})=\varepsilon_{p}-J(0)nm(m-\alpha)+h(m-\alpha).
\label{4.11'}
\end{equation}

To carry out the diagonalization of $\hat{\cal H}^{(2)}(m-1,m+1)$,
we introduce the creation and annihilation operators $\hat{b}_{{\bf
p}m+\sigma}$ ($\sigma=\pm 1$),
\begin{eqnarray}
\hat{a}_{{\bf p}m+\sigma}&=u_{m,\sigma}({\bf p})\hat{b}_{{\bf
p}m+\sigma}+v_{m,\sigma}({\bf
p})\hat{b}^{\dagger}_{-{\bf p}m-\sigma}, \nonumber  \\
\hat{a}^{\dagger}_{{\bf p}m+\sigma}&=u^{*}_{m,\sigma}({\bf
p})\hat{b}^{\dagger}_{{\bf p}m+\sigma}+v^{*}_{m,\sigma}({\bf
p})\hat{b}_{-{\bf p}m-\sigma}, \label {4.12}
\end{eqnarray}
in terms of which it has the diagonal form,
\begin{equation}
\hat{\cal H}^{(2)}(m-1,m+1)=\sum_{{\bf
p},\sigma}\omega_{m,\sigma}({\bf p})\hat{b}^{\dagger}_{{\bf
p}m+\sigma}\hat{b}_{{\bf p}m+\sigma}+E_{0}, \label{4.13}
\end{equation}
where $\omega_{m,\sigma}({\bf p})$ and $E_{0}$ are the excitation
spectrum and the ground state energy respectively. In order that the
introduced operators $\hat{b}^{\dagger}_{{\bf p}m+\sigma}$,
$\hat{b}_{{\bf p}m+\sigma}$ meet the canonical commutation
relations, the functions $u_{m,\sigma}({\bf p})$, $v_{m,\sigma}({\bf
p})$ must obey the relationships
\begin{eqnarray}
|u_{m,\sigma}({\bf p})|^{2}-|v_{m,\sigma}({\bf p})|^{2}=1, \nonumber \\
u_{m,\sigma}({\bf p})v_{m,-\sigma}(-{\bf p})-v_{m,\sigma}({\bf
p})u_{m,-\sigma}(-{\bf p})=0. \label{4.14}
\end{eqnarray}
Next, noting that
\[
\left[\hat{\cal H}^{(2)}(m-1,m+1),\hat{a}_{{\bf
p}m+\sigma}\right]=-\alpha_{m}({\bf p})\hat{a}^{\dagger}_{-{\bf
p}m-\sigma}-\gamma_{m,\sigma}\hat{a}_{{\bf p}m+\sigma}
\]
and expressing the right-hand side of this formula through
$\hat{b}_{{\bf p}m+\sigma}$, one gets
\begin{eqnarray}
\fl \left[\hat{\cal H}^{(2)}(m-1,m+1),\hat{a}_{{\bf
p}m+\sigma}\right]=&-\left(\alpha_{m}({\bf
p})v^{*}_{m,-\sigma}(-{\bf
p})+\gamma_{m,\sigma}u_{m,\sigma}({\bf p})\right)\hat{b}_{{\bf p}m+\sigma} \nonumber \\
&-\left(\alpha_{m}({\bf p})u^{*}_{m,-\sigma}(-{\bf
p})+\gamma_{m,\sigma}v_{m,\sigma}({\bf
p})\right)\hat{b}^{\dagger}_{-{\bf p}m-\sigma}, \label{4.15}
\end{eqnarray}
where $\gamma_{m,\sigma}$ is defined by
\begin{equation}
\gamma_{m,\sigma}=\varepsilon_{p}+\beta_{m\sigma}({\bf p})-\sigma h, \label{4.15'}
\end{equation}
moreover $\beta_{m\sigma}({\bf p})$ depends on the product
$m\sigma$. On the other hand, the straightforward use of
(\ref{4.13}), (\ref{4.12}) results in
\begin{eqnarray}
\fl \left[\hat{\cal H}^{(2)}(m-1,m+1),\hat{a}_{{\bf
p}m+\sigma}\right]&=v_{m,\sigma}({\bf p})\omega_{m,-\sigma}(-{\bf
p})\hat{b}^{\dagger}_{-{\bf p}m-\sigma}-u_{m,\sigma}({\bf
p})\omega_{m,\sigma}({\bf p})\hat{b}_{{\bf p}m+\sigma}.
\label{4.16}
\end{eqnarray}
The comparison of (\ref{4.15}) with (\ref{4.16}) gives the
coupled equations for $u_{m,\sigma}({\bf p})$ and
$v^{*}_{m,-\sigma}(-{\bf p})$,
\begin{eqnarray}
\left(\gamma_{m,\sigma}-\omega_{m,\sigma}({\bf
p})\right)u_{m,\sigma}({\bf p})+\alpha_{m}({\bf p})v^{*}_{m,-\sigma}(-{\bf p})=0,  \nonumber \\
\alpha_{m}({\bf p})u_{m,\sigma}({\bf
p})+\left(\gamma_{m,-\sigma}+\omega_{m,\sigma}({\bf
p})\right)v^{*}_{m,-\sigma}(-{\bf p})=0 \label{4.17}.
\end{eqnarray}
The condition for the existence of non-trivial solutions to the
coupled Eqs. (\ref{4.17}) along with the definitions
(\ref{4.9})-(\ref{4.11}), (\ref{4.15'}) result in the following
expression for the excitation spectrum:
\begin{eqnarray}
\fl \omega_{m,\sigma}({\bf p})=nm\sigma\left(J(0)-{1\over 2}J({\bf
p})\right) \nonumber \\
\pm\left[\varepsilon_{p}^{2}+\varepsilon_{p}nJ({\bf
p})(S(S+1)-m^{2})+\left({nJ({\bf p})m\over
2}\right)^{2}\right]^{1/2 }-\sigma h.
 \label{4.18}
\end{eqnarray}
At small ${\bf p}$ and $J(0)<0$ the obtained spectrum is real (the
state is stable) if the spin projections $m$ meet the inequality
$m^{2}>m_{c}^{2}$, where
$$
m_{c}=\left[\biggl({S(S+1)\over{1-nJ(0)/4\varepsilon_{p}}}\biggr)\right]^{1/2}, \quad m_{c} \leq S
$$
and the square brackets, as in (\ref{3.14}), are introduced to denote an integer part.

The functions $u_{m,\sigma}({\bf p})$, $v_{m,\sigma}({\bf p})$ are
found from (\ref{4.14}), (\ref{4.17}) and have the form
\begin{eqnarray*}
u_{m,\sigma}({\bf p})={\alpha_{m}({\bf
p})\over\sqrt{\alpha^{2}_{m} ({\bf
p})-\left(\omega_{m,\sigma}({\bf
p})-\gamma_{m,\sigma}\right)^{2}}}, \\
v_{m,\sigma}({\bf p})={{\omega_{m,\sigma}({\bf
p})-\gamma_{m,\sigma}}\over\sqrt{\alpha^{2}_{m}({\bf
p})-\left(\omega_{m,\sigma}({\bf
p})-\gamma_{m,\sigma}\right)^{2}}}.
\end{eqnarray*}
In fact, the functions $u_{m,\sigma}({\bf p})$, $v_{m,\sigma}({\bf
p})$ do not depend on $\sigma$ because, as it can be easily shown,
the quantity $\omega_{m,\sigma}({\bf p})-\gamma_{m,\sigma}$ is
independent of $\sigma$. The sign plus before the square root in
(\ref{4.18}) corresponds (for $\sigma=1$) to the wave, which
propagates in one direction, whereas the sign minus corresponds (for
$\sigma=-1$) to the wave propagating in opposite direction. Notice
that the obtained spectrum, as well as (\ref{4.11'}), contains only
the spin-exchange interaction amplitude and does not depend on the
potential interaction amplitude.

When $m=0$ (the antiferromagnetic ordering), the excitation
spectrum (\ref{4.18}) takes the form
\begin{equation*}
\omega({\bf p})=\sqrt{\varepsilon_{p}^{2}+\varepsilon_{p}nJ({\bf
p})S(S+1)}\pm h. \label{4.19}
\end{equation*}
In this case, for $h=0$ and $p\rightarrow 0$ we have
\[
\omega(p)=cp, \quad c=\sqrt{{n\over 2M}J(0)S(S+1)}.
\]
In ferromagnetic case (when $m=S$) the excitation spectrum is of
the form
\[
\omega({\bf p})=\varepsilon_{p}+{nJ({\bf p})S\over
2}(1-\sigma)+nJ(0)S\sigma-\sigma h.
\]

The similar mathematical manipulations with $\hat{\cal
H}^{(2)}(m)$ lead to another mode of excitation spectrum, which
depends  both on potential and spin-exchange interaction amplitudes,
\begin{equation}
\omega_{m}({\bf
p})=\sqrt{\varepsilon_{p}^{2}+2\varepsilon_{p}n(U({\bf
p})+m^{2}J({\bf p}))}. \label{4.20}
\end{equation}
The stability region for this spectrum at small ${\bf p}$ is given by (\ref{3.14}).

The functions $u_{m}({\bf p})$ and $v_{m}({\bf p})$ can be found
immediately,
\begin{eqnarray*}
u_{m}({\bf p})={{\varepsilon_{p}+\omega_{m}({\bf p})}\over
2\sqrt{\varepsilon_{p}\omega_{m}({\bf p})}}, \quad v_{m}({\bf
p})={{\omega_{m}({\bf p})-\varepsilon_{p}}\over
2\sqrt{\varepsilon_{p}\omega_{m}({\bf p})}}.
\end{eqnarray*}

When $J({\bf p})=0$ the excitation spectrum (\ref{4.20}) coincides
with the spectrum found by Bogoliubov \cite{Bogoliubov}. At small
${\bf p}$, the spectrum has the following phonon behavior:
\[
\omega_{m}({\bf p})=cp, \quad c=\sqrt{{n\over M}(U(0)+m^{2}J(0))}.
\]
In this formula, as well as in (\ref{4.20}), we have chosen the
arithmetic value of the square root.

In conclusion, we have studied BEC of atoms with arbitrary spin in a
magnetic field on the basis of the model for weakly interacting Bose
gas. We have derived the equation which describes the ground state
of spin-$S$ BEC at temperature $T\to 0$ and found its particular
solution. This solution corresponds to the formation of BEC of
spin-$S$ atoms with a definite spin projection $m$ onto direction of
a magnetic field that is also true for an ideal Bose gas
\cite{Akhiezer}. The explicit expression for thermodynamic potential
being a function of chemical potential and spin projection has been
obtained. It generalizes the thermodynamic potential for weakly
interacting Bose gas to the case when both potential and
spin-exchange interactions act between bosons. The thermodynamic
stability of the state under consideration has been studied and the
spin projections, which give a minimum of thermodynamic potential,
have been found. These projections are given by the integer part of
the ratio of potential to spin-exchange interaction amplitude. The
expressions for low-lying collective modes related to the solution
(\ref{3.9}) have been obtained. Notice that Eq. (\ref{3.8}) for
order parameter has also other solutions different from (\ref{3.9}).
Our present research deals with seeking such solutions.

\ack
A. Peletminskii would like to thank the Abdus Salam ICTP
(Trieste, Italy) for the support, warm hospitality, and
stimulating research environment during the visit to the Centre
within the framework of Associateship Scheme in summer 2005.
He is also grateful to Professor S. Shenoy for his attention to
the work.

\end{document}